\documentclass[twocolumn,nofootinbib,amsmath,amssymb,a4paper]{revtex4}

\usepackage{graphicx}
\usepackage{dcolumn}
\usepackage{bm}

\begin{document}

\title{Measuring $\gamma$ with $B^0 \rightarrow D^0 K^{*0}$ at BaBar}

\author{S. Pruvot$^{*}$, M.H. Schune$^{*}$, V. Sordini$^{*,o}$, A. Stocchi}
 \email{viola.sordini@slac.stanford.edu}
\affiliation{$^*$Laboratoire de l'Acc\'el\'erateur Lin\'eaire, Orsay, France\\
 $^o$Universit\`a di Roma ``La Sapienza" \& INFN Roma I, Italy \\
}

\begin{abstract}
We present a feasibility study for a new analysis for extracting the angle $\gamma$ 
of the Unitarity Triangle from the study of the neutral B meson decays. 
We reconstruct the decay channel 
$B^0\rightarrow \bar{D^0} K^{*0}$ with the $K^{*0}\rightarrow K^{+}\pi^{-}$ 
and the $\bar{D}^{0}\rightarrow K_{S}\pi^{+}\pi^{-}$ using a $D^0$ Dalitz 
analysis technique. 
The sensitivity to the angle $\gamma$ comes from the interference 
of the $b \rightarrow c$ and $b \rightarrow u$ processes contributing to 
the same final state and by the fact that the $B^0 (\bar{B^0})$ can be 
unambiguosly identified through the sign of electric charge of the kaon from 
$K^{*0} (\bar{K^{*0}})$ decay.
The impact of the result of such analysis is evaluated for the actual 
BaBar statistics.
\end{abstract}

\maketitle

\section{The technique}

Various methods related to $B^- \to D^{(*)0}K^{(*)-}$ decays have been proposed to
determine the UT angle $\gamma$. Within them, the one that gives the best 
error on $\gamma$ is the Dalitz method \cite{DalitzBaBar}\cite{DalitzBelle}.

We propose to measure the angle $\gamma$ using a Dalitz
analysis applied to neutral B mesons. In general, since the neutral B mesons mix, 
interference effects between $b \to c$ and $b \to u$ decay amplitudes in $B^0$ 
decays (for instance into $D^{(*)\pm}\pi^\mp$ final states) are studied for the 
determination of the combination of UT angles $2\beta+\gamma$.
In this case the tagging technique and a time dependent analysis are required.
This can be avoided if the final states contain a particle which allows to
unambiguously identify if a $B^0$ or $\bar{B}^0$ has decayed. This is the case 
of neutral B mesons decaying into $\bar{D}^0  [K^+ \pi^-]$ final states, where 
the flavor of the B meson can be determined through the sign of the electric 
charge of the Kaon. 

One of the relevant parameter in those kind of analysis is the ratio between 
the $b \to u$ and $b \to c$  amplitudes, which can be expressed by :

\begin{equation}
r_B (D^0K^{*0}) = \frac{|{\cal A}({D}^{0} {K}^{*0})|}{|{\cal A}(\bar{D^{0}} {K}^{*0})|}    
                = \frac{|V_{ub}|}{\lambda |V_{cb}|}
                \frac{{\cal A_{\rm{strong}}}({D}^{0} {K}^{*0})}{{\cal A_{\rm{strong}}}(\bar{D^{0}} {K}^{*0})} .
\label{eq:rb}
\end{equation}

The sensitivity to $\gamma$ is proportional to the $r_B$ value.
Considering the CKM factors in the $b \to c$ and $b \to u$ transitions and 
the fact that in both cases the processes are mediated by a color suppressed 
diagram, we expect $r_B (D^0K^{*0})$ to be in the range [0.3-0.5], larger than 
for the equivalent ratio in the charged B sector, which has been found to 
be of the order 0.1~\cite{Bona:2005vz}.

\section{Structure of the analysis}

In the analysis on data we would perform an unbinned maximum likelihood 
fit in order to extract the interesting quantities: the likelihood will 
contain a product of a yields pdf and a CP pdf (only the latter has, within 
its variables, the weak phase $\gamma$).

In our feasibility study, we assume the yields pdf to be a function 
of the classic BaBar analysis variable $m_{ES}$ \footnote{the beam-energy 
substituted mass $m_{ES}  \equiv \sqrt{(\sqrt{s}/2)^2-{p^*_B}^2}$, where 
the asterisk denotes evaluation in the $\Upsilon (4\mbox{s})$ CM frame} 
and of a variable that is able to discriminate between $B\bar{B}$ and 
continuum events ($q\bar{q}$ events, with $q=u, d,s, c$).
For the distributions of those variables we make some realistic 
assumptions inspired by BaBar analysis.

In writing the CP pdf, we have an additional difficulty, with respect to the 
$B^{-} \rightarrow D^0 K^{-}$ Dalitz analysis. 
This is due to the fact that the natural width of the $K^{*0}$ is not small 
($\sim$50~\mbox{MeV}) and hence the interference with the other 
$B^0\to \bar{D^0} (K\pi)_{\rm{non-}K^*}^0$ processes may not be negligible. 
We follow the formalism introduced in~\cite{gronau2002} and also used in the 
$B^- \rightarrow D^0 K^{*-}$ BaBar analysis in order to solve this problem.

We introduce the effective quantities $r_S$, $k$ and $\delta_S$ defined as :
\begin{eqnarray}
&& r_s^2=\frac{\Gamma(\bar{B^0}\to \bar{D^0} (K^+\pi^-))}{\Gamma(\bar{B^0}\to D^0 (K^+\pi^-))}
=\frac{\int dp\ |A_{up}^2|}{\int dp\ |A_{cp}^2|}\label{eq:rs_square}\\
&& ke^{i\delta_S}=\frac{\int dp\ A_{cp}A_{up}e^{i\delta_p}}{\sqrt{\int dp\ |A_{cp}^2|\ \int dp\ |A_{up}^2|}}\, ,\label{eq:k}
\end{eqnarray}
where $0\leq k \leq 1$ for the Schwartz inequality and $\delta_S\in[0,2\pi]$.
In the limit of a $B\to$2-body decay, such as $\bar{B^0}\to D \bar{K^0}$ 
we have : $r_S \rightarrow r_B$, $\delta_S \rightarrow \delta_B$ (the phase difference between 
the $V_{ub}$ and the $V_{cb}$ amplitudes of Eq.\ref{eq:rb}) and k=1.
$A_{cp}$, $A_{up}$ are real and positive. The subscript $c$ and $u$ refer to the $b\to c$ and 
$b\to u$ transitions, respectively. The index $p$ indicates the position in the phase space of 
$D^0 K^+ \pi^-$.

Applying this formalism, we write the CP pdf as follows:
\begin{eqnarray}
\Gamma(\bar{B^0}(B^0)\to D[\to K^0_S\pi^-\pi^+]K^-\pi^+(K^+\pi^-))  \propto  \nonumber \\
|f_\mp|^2+r_s^2|f_\pm|^2+ 2kr_s|f_\mp||f_\pm|\cos(\delta_S+\delta_D(m_\mp^2,m_\pm^2)\mp\gamma)\, ,
\label{eq:dalitzrate1}
\end{eqnarray}
where $\delta_D(m_\mp^2,m_\pm^2)$ is the strong phase difference between 
$f(m_\pm^2,m_\mp^2)$ and $f(m_\mp^2,m_\pm^2)$ and $r_s$, $k$ and $\delta_S$ 
are defined in Eqs.~(\ref{eq:rs_square}) and~(\ref{eq:k}). We have simplified the notation using
$f_{\pm} \equiv f(m_\pm^2,m_\mp^2)$ and $f_{\mp} \equiv f(m_\mp^2,m_\pm^2)$.

The Dalitz structure of the decay $D^0\rightarrow K_{S}\pi\pi$ is well known 
and it has already been used in Dalitz analysis in the charged B sector 
\cite{DalitzBaBar} \cite{DalitzBelle}.

\section{the $k$ and $r_S$ parameters}

We performe a study to evaluate the possible variation of $r_S$ and $k$ on the B 
Dalitz plot. For this purpose we use a B Dalitz model as suggested by the recent 
measurements \cite{ref:polci}, \cite{ref:cavoto}. The  model assumed for the decay 
parametrizes the amplitude $A$ at each point $k$ of the Dalitz plot as a sum of 
two-body decay matrix elements and a non-resonant term according to the following 
expression :
\begin{equation}
A_{c_k(u_k)}e^{i\delta_{c_k(u_k)}}=\sum_j a_je^{i\delta_j} BW^j_k(m,\Gamma,s) + a_{nr} e^{i \phi_{nr}}
\label{eqres}
\end{equation}
We consider a region within $\pm$ 48 MeV from the nominal mass of the $K^{*0}$(892) 
resonance and we obtain the distribution of $r_S$, $k$ and $k \times r_S$ by randomly varying 
all the strong phases between [0-2$\pi$] and the amplitudes up to $\pm 30\%$
of their nominal value (shown in table \ref{tab:b0}). 
Only for the $D_{s,2}(2573)^{\pm}$ resonance, the variation is up to $\pm 200\%$ of its 
nominal value.
\begin{table}
\begin{center}
\begin{tabular}{l c c c c c} \hline \hline
                        & $M (GeV/c^2)$ & $W (GeV/c^2)$ & $J^P$ & $a(V_{cb})$   & $a(V_{ub})$  \\ \hline
 $D_{s,2}(2573)^{\pm}$  &    2.572         &      0.015        & $2^+$ &   0           & 0.02         \\ 
 $D_{2}^*(2460)^{\pm}$  &    2.459         &      0.029        & $2^+$ &  1.0          &  0           \\
 $D_{0}^*(2308)^{\pm}$  &    2.403         &      0.283        & $0^+$ &  1.0          &  0           \\
 $D_{0}^*(2010)^{\pm}$  &    2.0100        &      0.000096     & $1^-$ &  -            &  -           \\
 $K^*(892)^{0}$         &    0.89610       &      0.0507       & $1^-$ &  1.0          & 0.4          \\
 $K_{0}^*(1430)^{0}$    &    1.412         &      0.294        & $0^+$ &  0.3          & 0.12         \\
 $K_{2}^*(1430)^{0}$    &    1.4324        &      0.109        & $2^+$ &  0.15         & 0.06         \\
 $K^*(1680)^{0}$        &    1.717         &      0.322        & $1^-$ &  0.2          & 0.08         \\
 Non resonant           &       -          &         -         &  -    &  -            &  - \\ \hline \hline
\end{tabular}
\end{center}
\caption{List of mass ($M$), widths ($W$) and quantum numbers  of the resonances considered in our model. 
The last two columns present the chosen values of the coefficients $a_j$ for the $V_{cb}$ and $V_{ub}$ 
transitions respectively. Note that the phase $\delta_j$ are not indicated and their choice is arbitrary.}
\label{tab:b0}
\end{table}

\begin{figure}[!tbp]
\begin{center}
\includegraphics[width=0.22\textwidth]{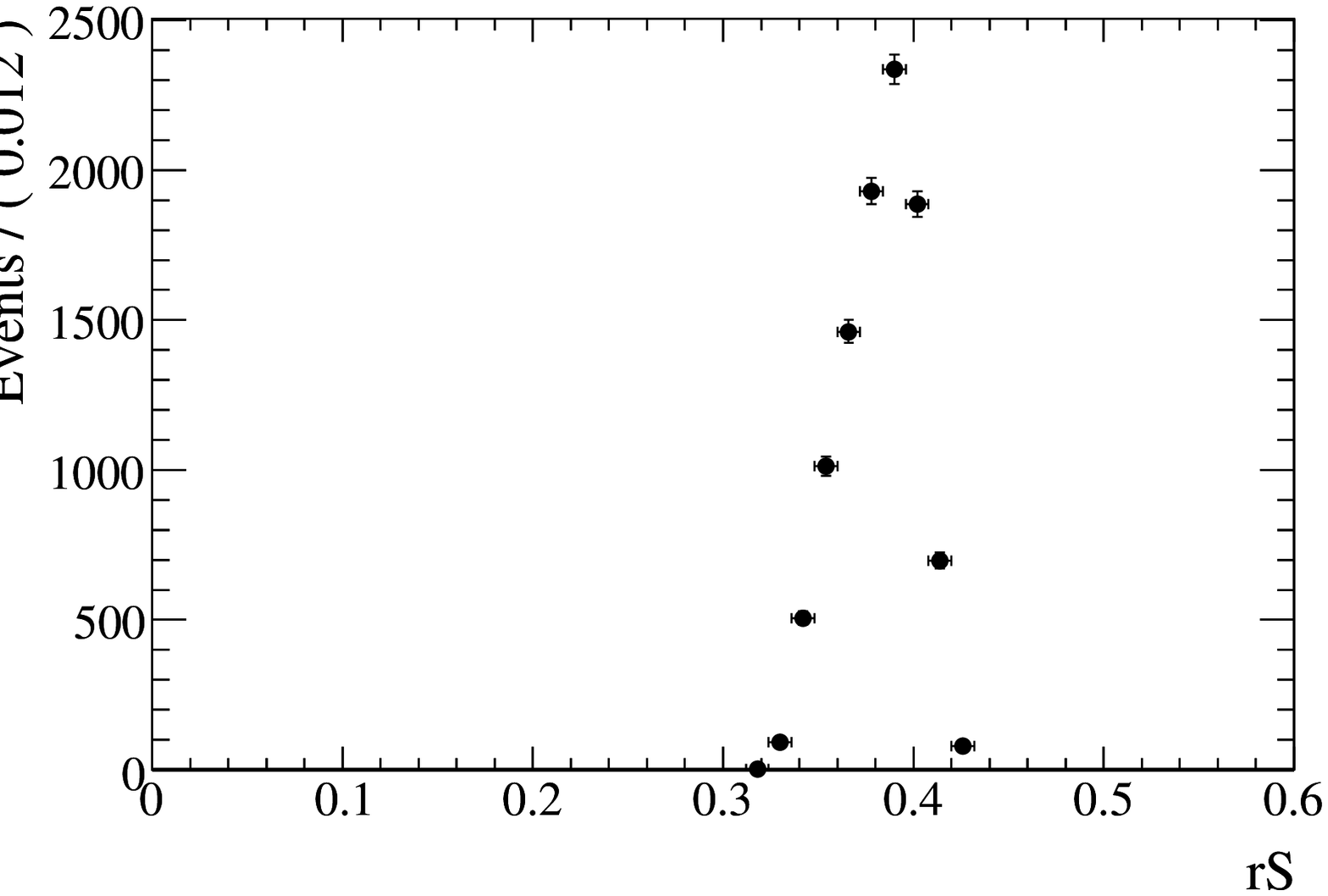} 
\includegraphics[width=0.22\textwidth]{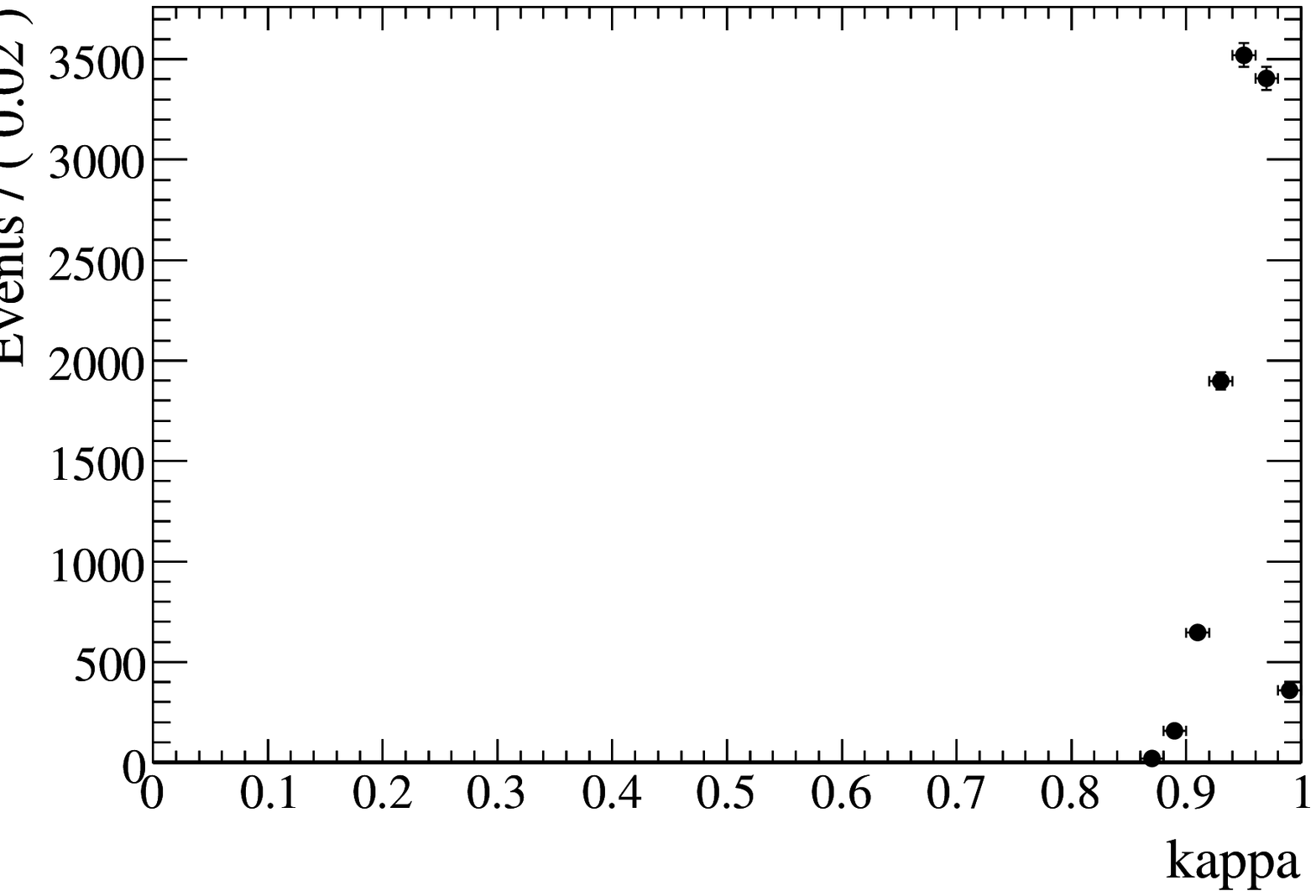}
\end{center}
\caption{\it {Distribution of $r_S$ and $k$ in a region within $\pm$ 50 MeV the nominal mass of the $K^{*0}$(892) 
resonance ($m_{K \pi}$ in the range [0.7159-0.8951] $GeV^2$). This distribution has been obtained by randomly 
varying all the strong phases (between [0-2$\pi$]), the amplitudes 
(between [0.7-1.3] of their nominal value (except those relative to $K^*$ which has been fixed to 0.4 and for 
the $V_{ub}$ amplitude of the $D_{s,2}(2573)^\pm$ which has been varied between [0.-2.] of its nominal value) 
given in Table \ref{tab:b0}).}}
\label{fig:rs}
\end{figure}

The distribution we obtain for $r_S$ and $k$ are shown in figure \ref{fig:rs};

The results is that, in $K^{*0}$(892) mass region, $r_S$ can vary between 
0.3 and 0.45 depending upon the values of the phases and of the amplitudes 
contributing in the $K^{*0}$ region. In absence of pollution we would 
have expected $r_S=r_B=0.4$.  The distribution of $k$ is quite narrow, 
centered to 0.95 with an error of 0.03.
 
For this reason the value of $k$ can be fixed to 0.95 and the effect of 
its variation (within a reasonable interval) will be considered a systematic error. 

\section{CP fit studies}

We perform an intensive toy-MC study on the CP fit, assuming the actual BaBar 
statistics ($350 \mbox{fb}^{-1}$). We assume to have $35$ signal events, about 
$2000$ continuum background events and $250$ $B\bar{B}$ background events.

\subsection{Polar coordinates}
\begin{figure}[!ht]
\begin{center}   
\includegraphics[width=0.2\textwidth]{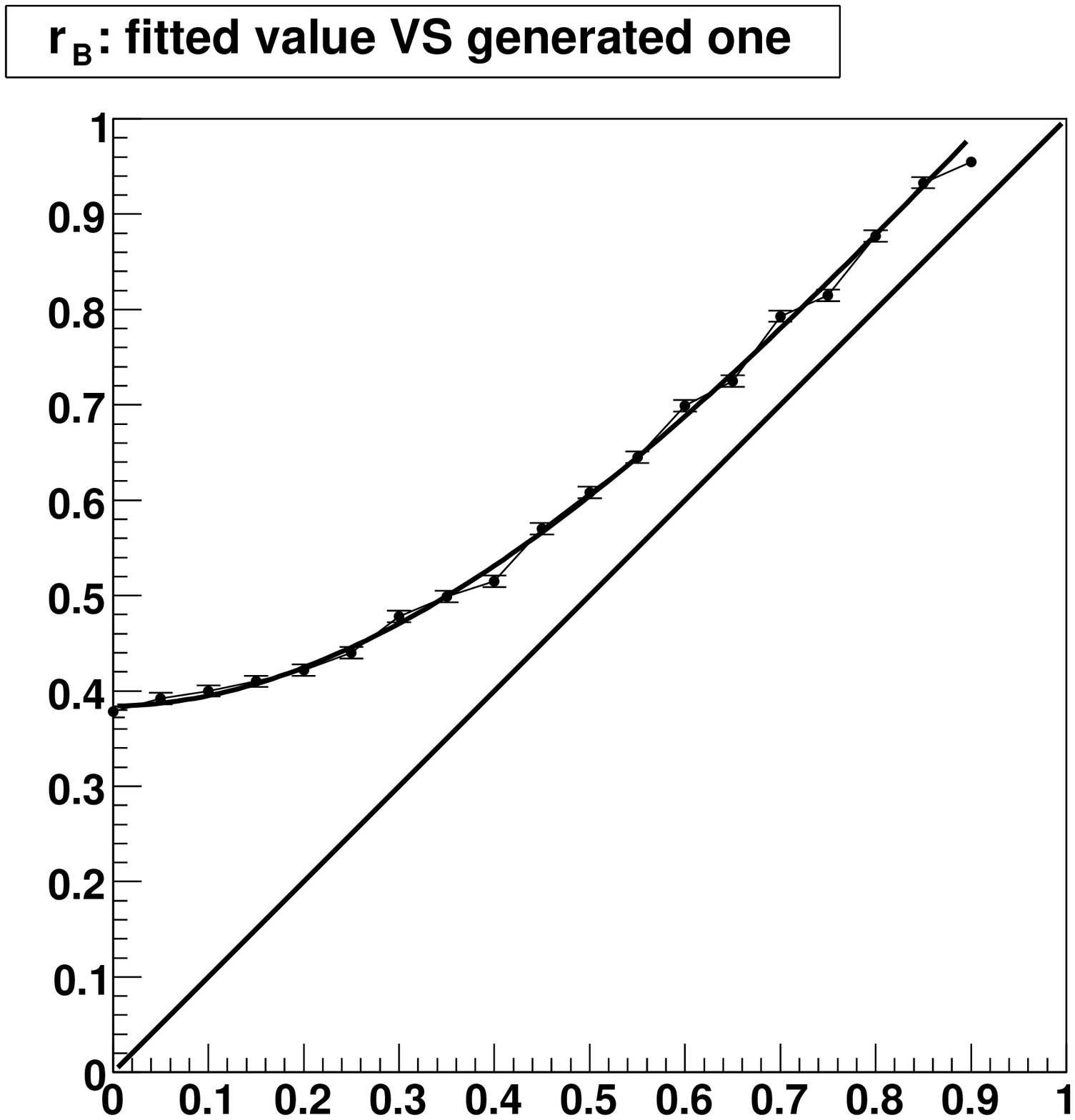}
\includegraphics[width=0.2\textwidth]{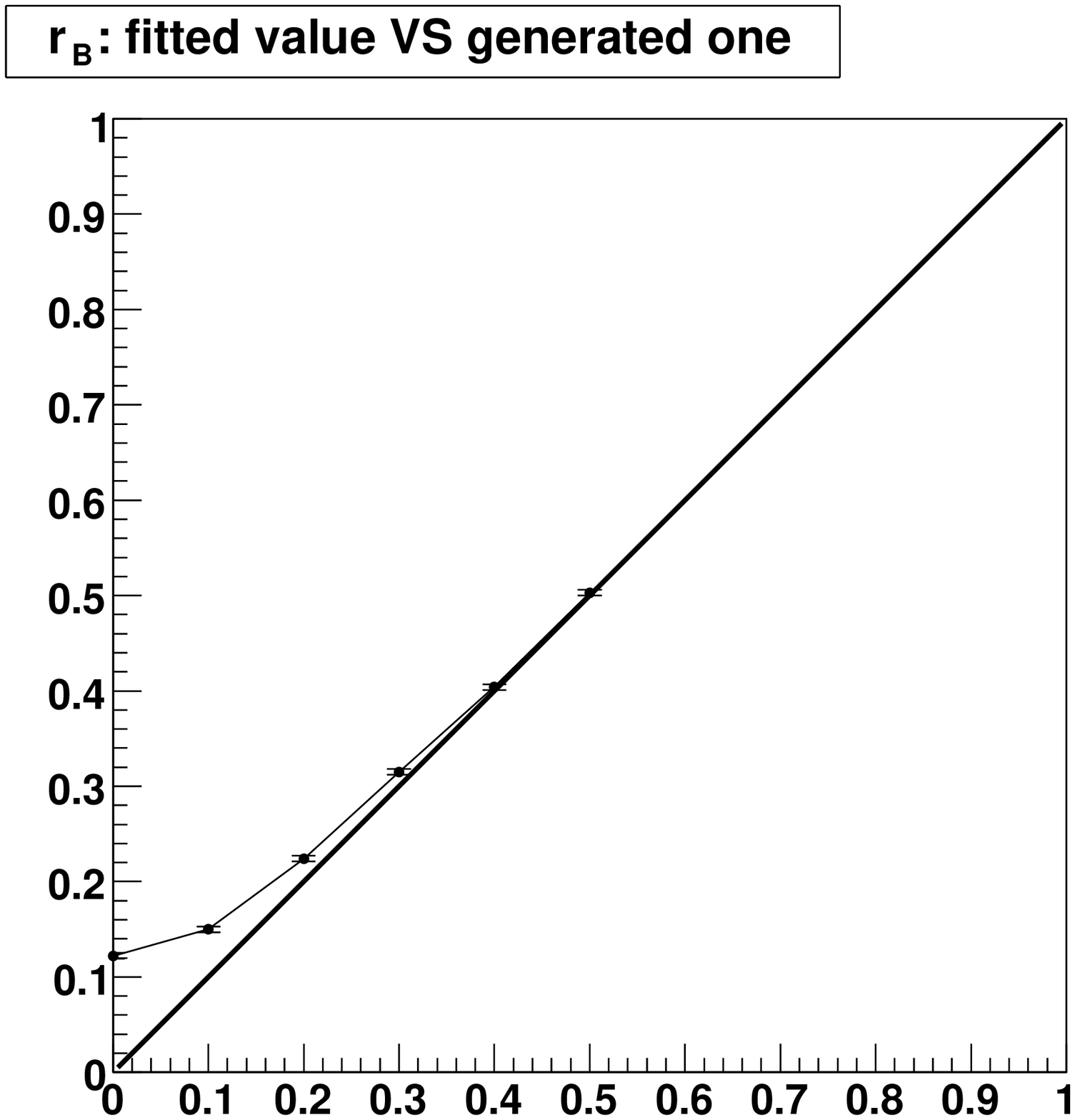}
\caption{Distribution of $r_{S}^{FIT}$ vs $r_{S}^{GEN}$ from toy MC for 
different generated values.The toy-MC have been generated assuming 
$350  \mbox{fb}^{-1}$(left plot) and 10 
times the statistics (right plot).\label{fig:Linearity}}
\end{center}
\end{figure}

We first perform toy-MC studies in polar coordinates.
In figure \ref{fig:Linearity} we observe the known linearity problem: 
due to the dependence of the likelihood on $r_S$, we tend to get from the 
fit a value of $r_S$ higher than the true (generated) one and consequently to 
underestimate the error on $\gamma$.

It is evident that the linearity effect (still visible in the results of the 
tests at high statistics) is not the only problem that affects our measurement.
In fact, as it can be seen from the left plot, also for high values of $r_{S}^{GEN}$
the results for $r_{S}^{fit}$ vs $r_S^{GEN}$ do not tend to the curve $r_{S}^{fit}=r_S^{GEN}$ 
(the black curve in the plots).
We thus conclude that we cannot fit in polar coordinates 
(with $r_S$, $\gamma$ and $\delta_S$ floating) because of the two effects:
the linearity and the low statistics and that is the second one that dominates 
(at least for $r_{S}\approx 0.4$, as it is expected to be for our measurement).

\subsection{Cartesian coordinates}
We then perform a toy-MC study in cartesian coordinates \cite{ref:cart}.
The use of these coordinates normally solves the linearity 
problem (\cite{DalitzBaBar}, \cite{DalitzBelle}).

From the toy-MC the four variables appear to be biased and show a non-Gaussian behaviour. 
In table \ref{tab:cartcoordpull}, we summarize the results of toy-MC 
where we generated for the yields the 
values we expect on $350 fb^{-1}$ (left column) and the results of toy-MC for 
10 times the now available BaBar statistics (right column).
As it can be seen in the left column, all the variables are biased and 
the $\sigma$ of their pulls are not compatibles with $1$. This effect disappears at high statistics. 
We conclude that, with the now available statistics, we cannot perform the 
measurement in cartesian coordinates.

\begin{table}[htpb]
\begin{center}
\begin{tabular}{c c c }
\hline \hline
 - & $now$ & $10 x$ stat\\ \hline
$\mu^{PULL}_{x_{+}}$    & $-0.52 \pm 0.05$  &  $-0.04 \pm 0.05$      \\
$\sigma^{PULL}_{x_{+}}$ & $ 0.82 \pm 0.03$  &  $ 0.97 \pm 0.04$    \\ 
$\mu^{PULL}_{x_{-}}$    & $-0.07 \pm 0.05$  &  $-0.02 \pm 0.05$   \\ 
$\sigma^{PULL}_{x_{-}}$ & $ 0.78 \pm 0.04$  &  $ 0.99 \pm 0.04$   \\ 
$\mu^{PULL}_{y_{+}}$    & $-0.18 \pm 0.05$  &  $-0.05 \pm 0.06$   \\ 
$\sigma^{PULL}_{y_{+}}$ & $ 0.79 \pm 0.04$  &  $ 1.01 \pm 0.04$    \\
$\mu^{PULL}_{y_{-}}$    & $ 0.40 \pm 0.05$  &  $-0.03 \pm 0.05$    \\
$\sigma^{PULL}_{y_{-}}$ & $ 0.79 \pm 0.04$  &  $ 1.03 \pm 0.04$   \\  \hline\hline
\end{tabular}
\end{center}
\caption{Pull distributions for {\bf cartesian} coordinates on 500 toy-MC with a 
signal background configuration similar to the one we find on data (left column) and 
for 10 times the statistics (right column).\label{tab:cartcoordpull}}
\end{table}

\subsection{Measurement strategy}

We conclude that, with the now available BaBar statistics, the measurement 
strategy would be extracting from the fit $\gamma$ as a scan with 
respect to $r_B$ (i.e. by performing a likelihood scan on $\gamma$ for each value of $r_B$ 
with $\gamma$, $\delta$ and all the yields parameters floated in the fit).

In this way we would extract from data the maximum possible information 
for the available statistics: an information on $\gamma$ and not on $r_B$.
The $\gamma$ distribution obtained in that way, already very interesting on its own, would be 
very precious when combined with an experimental input for $r_B$.
The coverage tests on toy-MC generated with different values of $r_S$ 
show that we have no bias in this fit configuration.

\section{impact of the measurement}
\begin{figure}[!htbp]
\begin{center}   
\includegraphics[width=0.4\textwidth]{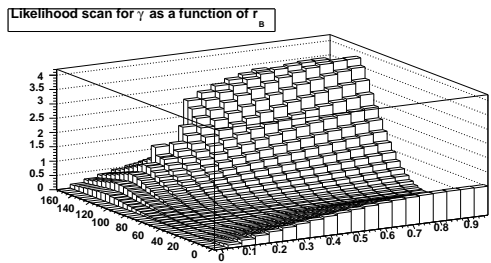}
\caption{Negative Log Likelihood scan for $\gamma$ for different values of $r_B$. 
The distribution is obtained on a chosen toy-MC.\label{fig:gammaScan_rBfix_toy}}
\end{center}
\end{figure}

In this section we show, on one chosen toy-MC, what would be the 
impact of a real mesurement.
In figure \ref{fig:gammaScan_rBfix_toy} we show the output of our fit: a 
likelihood scan of $\gamma$ with respect to $r_S$.

This distribution, when combined with a fake $r_S$ measurement ($r_S = 0.35\pm 0.15$) 
gives $\gamma$ with an error of $46^{o}$.
A measurement of $r_S$ with such an error is what we could expect to obtain, if 
$r_S=0.35$, on $500 \mbox{fb}^{-1}$ data from an analysis of $B^0 \rightarrow D^0 K^{*0}$ 
with the $D^0$ in flavor modes.

Clearly, this is just a example on a chosen toy-MC and it is not necessarily 
representative of what we will find on data.
Indeed we found, on toy-MC studies, that in a small percentage of cases 
we can be much less sensitive to $\gamma$. 
That problem has been found to depend on the low number of signal events 
(such that in some cases those events happen to be in poorly sensitive regions of the 
$D^0$ Dalitz plot) and it will therefore disappear with the increasing of 
the statistics.
Still, our test shows that we could have a result on $\gamma$ that is competitive 
with the charged B Dalitz analysis (the analysis made on BaBar data gives today 
an error on $\gamma$ of $42^{o}$).


\begin{thebibliography}{99}
\bibitem{ref:cart}The parameterization given in Eq.~(\ref{eq:dalitzrate1}) can be rewritten in terms of the 
cartesian coordinates $x_{s\pm} =  Re[kr_s e^{i(\delta_s\pm\gamma)}]$ and 
$y_{s\pm} = Im[kr_s e^{i(\delta_s\pm\gamma)}]$ as 
$\Gamma(\bar{B^0}(B^0)\to D[\to K^0_S\pi^-\pi^+]\bar{X_s^0}(X_s^0)) \propto$  \\
$|f_\mp|^2+r_s^2|f_\pm|^2+2 \left[ x_{s\mp} Re[f_{\mp}f_{\pm}^*] + y_{s\mp} Im[f_{\mp}f_{\pm}^*] \right]$
\end{thebibliography}
\end{document}